\documentstyle[12pt,aaspp4]{article}
\tighten
\begin{document}

\title{Gold alignment \& internal dissipation}
\author{A. Lazarian\altaffilmark{1}}
\affil{Astronomy Department, University of Texas, Austin,
    TX 78712-1083}

\altaffiltext{1}{Present address:
Department of Astrophysical Sciences, Princeton University,
Princeton, NJ 08544}

\begin{abstract}
The measures of mechanical alignment were obtained for both prolate and 
oblate grains when their temperature is comparable with grain kinetic 
energy devided by $k$, the Boltzmann constant. For such grains, the 
alignment of angular momentum, ${\bf J}$, with the axis of maximal inertia, 
$\bf a$, is only partial. This substantially alters the alignment as 
compared with the results in Lazarian (1995) and Roberge, Hanany \& 
Messinger (1996) obtained on the assumption of perfect alignment.
 We also describe the Gold alignment when the Barnett dissipation is
suppressed and derive an analytical expression which relates
the measure of alignment  with parameters of grain nonsphericity and
the direction of the gas - grain drift. This solution
provides the lower limit for the alignment measure, while the upper
limit is given by the analytics derived in Lazarian (1994). Using results
of a recent study of incomplete internal relaxation in Lazarian \&
Roberge (1996), we find measures of alignment for the whole range of 
ratios of grain rotational energy to $k$ over $T_{\rm s}$, where
 $T_{\rm s}$ is the
grain temperature. To describe alignment for mildly supersonic drifts, we
suggest an analytical approach which provides good correspondence with
the results of direct numerical simulations in Roberge, Hanany \& Messinger 
(1995). We also extend our approach to account for the simultaneous action 
of the Gold and Davis-Greenstein mechanisms.
\end{abstract}


\section{Introduction}

Understanding  the observed alignment of the ISM grains is yet 
unsolved astrophysical problem (see
Hildebrand 1988, Whittet 1992, Goodman et al. 1995, Roberge 1996). This 
limits the use of polarimetry data for studying interstellar magnetic fields.

The mechanism of mechanical alignment of thermally rotating grains 
was pioneered by Gold (1951) the same year that the classical 
paper introducing the paramagnetic alignment
 by Davis \& Greenstein (1951) was published. Originally,
Gold suggested that grain alignment arises from 
cloud - cloud collisions (Gold 1951, 1952), but it was shown in 
Davis (1955), that such collisions can align only an insignificant 
fraction of interstellar grains. Therefore a further study of the mechanism 
was devoted mainly to the alignment in the vicinity of bright sources, where 
radiation
pressure can drive grains to supersonic velocities (see Purcell 1969, Aitken
et al. 1995). A new important idea, 
put forward by Roberge \& Hanany (1990), that grains can be aligned
by ambipolar diffusion made the Gold alignment  more promising (see also
Roberge, Hanany \&  Messinger 1995 hereafter RHM). 
Our study in Lazarian (1994, henceforth Paper I, and 1995a) 
showed that the role of mechanical processes had been
underestimated. It was found that pervasive MHD
waves\footnote{Here and further on
 in the paper when speaking about  MHD waves we mean both Alfv\'{e}nic and 
magnetosonic waves.} can produce grain alignment even within ideal MHD.

The goal of this paper is to provide an analytical quantitative description of 
the mechanical alignment of thermally rotating grains, refered to as
``Gold alignment''. We distinguish between the  Gold  
and mechanical alignment of suprathermally rotating grains (see Lazarian
1995a,b, Lazarian \& Efroimsky 1996, Lazarian, Efroimsky \& Ozik 1996).

Although the expressions for the measure of alignment corresponding
to  Gold alignment were obtained in Paper I, the 
shortcoming of this study was that only 
 perfect alignment of angular momentum, ${\bf J}$, with the 
 the principal axis of the maximal moment of inertia, ${\bf a}$, 
(henceforth the axis of  major inertia) was considered.
Although this regime is  valid for highly supersonic motions 
discussed in Lazarian (Paper I), the wealth of the ISM conditions 
presents us with a wide range of other options.

It is well known from theoretical mechanics, that internal dissipation 
of energy cannot change grain angular momentum and, for a grain with 
fixed angular momentum, the minimal energy corresponds to
 rotation about the axis of  major inertia. For suprathermally rotating 
grains, efficient internal dissipation of energy leading to nearly 
perfect alignment of angular momentum with the grain axis of major inertia 
was discovered by Purcell (1979) (see
also Spitzer \& McGlynn 1979). For thermally rotating grains, the alignment
is only partial (Paper I, Lazarian \& Roberge 1997, henceforth LR97).

To relate polarimetry observations, that are influenced by
the alignment of the long grain axis, and the theory, that deals with
the alignment of angular momentum, $\bf J$, one has 
  to describe the alignment of $\bf J$ not only in respect to magnetic
field\footnote{Magnetic field acts as the axis of alignment because the
time-scale of Larmor precession is usually much shorter than the time-scale 
of alignment. We remind the reader that the rapid Larmor precession
arises from magnetic moments of  grains; those moments are caused by
the Barnett effect (Dolginov \& Mytrophanov 1976).}, but also 
in respect to grain axes. These  can be called
external and internal alignment, respectively. 
Internal alignment arises both from the difference
in grain moments of inertia (``Maxwellian alignment'') and
due to  the Barnett relaxation\footnote{As a rule, the Barnett relaxation
dominates internal relaxation (see Purcell 1979).} (``Barnett alignment'').

Below we use the statistical approach introduced in LR97. The gist 
of it is to describe deviations of $\bf J$ from the
axis of major inertia, $\bf a$, as thermal fluctuations from the position of
static equilibrium  when the time of internal relaxation is much 
shorter than that of the gaseous damping. By introducing the  grain rotational
temperature, which depends on 
both the temperature of the ambient gas and on the drift velocity, we
show that, when the ratio of this temperature to the temperature 
of grain material tends to infinity, $\bf J$ is coupled with $\bf a$, and this
justifies our approach  in Paper I. For finite ratios however, 
the deviations of $\bf J$ become important for the alignment.

To describe Gold alignment for mildly supersonic velocities, 
we have to account for the rms motions of gas atoms. To do this, we assume
that grains are subjected to a superposition of fluxes within 
a small angle to the direction of the initial flux and the value of this 
angle depends on the ratio of the flux velocity to the mean velocity of 
 thermal motions. This enables us to obtain an asymptotic solution 
which provides  fair agreement with the numerical
calculations in RHM for the case of perfect coupling between $\bf J$ and 
$\bf a$. Within the same model of perfect coupling the agreement with
numerics in RHM is  obtained for the joint action of
Gold and Davis-Greenstein mechanisms. A comparison between numerics and 
analytics for the general case of incomplete internal alignment is
planned, and we  anxiously follow the progress of numerics in this
direction.

The structure of the paper is as follows. 
First of all, we remind our reader of the concept of internal
relaxation  and describe how grain drift influences internal alignment 
(Sect.~2). Then in Sect.~3, we obtain analytical solutions for the 
alignment measure when internal relaxation is negligible and 
compare these results with the other extreme case, namely, when $\bf J$ is 
perfectly coupled with $\bf a$.
The treatment of the problem for an arbitrary degree of internal alignment
is given in Sect.~4. In Sect.~5, we compare our results with numerics and 
argue that our analytical approach 
provides an adequate description of the alignment both for mildly supersonic
velocities and for Gold and Davis-Greenstein processes acting together. 
A short summary of results is presented in Sect.~6.

\section{Internal alignment}

As pointed out above, to relate theory and observations it is essential to 
describe the 
alignment of grain long axes in respect to magnetic field\footnote{In a special
case, e.g. in the atmospheres of comets, the alignment can happen on
the time-scale less than that of precession. In such situations, the
alignment should be defined in respect to the flow, which is 
similar to the alignment when the flow is
directed along magnetic field and therefore we do not discuss it 
separately.}. This alignment is determined uniquely by the alignment of 
grain angular momentum  only if the angular
momentum is coupled with the axis of major inertia, $\bf a$. Further we study a
general problem when the internal alignment is partial.

Theoretical studies deal with the measure of 
alignment of angular momentum $\bf J$\footnote{Our notations are
different from the ones used in Paper I. In an attempt to insure 
that the system of notations used by different authors is universal 
we adopt the system of notations suggested in review by Roberge (1996).}
\begin{equation}
Q_{\rm J}=\frac{3}{2}\langle \cos^{2}\theta_1 - \frac{1}{3} \rangle ~~~,
\label{j1}
\end{equation}
where $\theta_1$ is the angle between ${\bf J}$ and the direction of 
magnetic field,  while polarimetry provides us with the data on 
the Rayleigh reduction factor (Greenberg 1968)
\begin{equation}
R=\frac{3}{2}\langle\cos^{2}\hat{\theta}\rangle-\frac{1}{2} ~~~,
\label{2.45}
\end{equation}
where $\hat{\theta}$ is the angle between the direction of magnetic field
and the symmetry axis of a spheroid approximating the grain.  
The question how to relate these two quantities 
was on the astrophysical agenda from the very beginning of research 
in the area. First in Jones \& Spitzer (1967), it was assumed that the 
distribution function of $\bf J$ in the grain reference frame 
is independent of the alignment of $\bf J$ in respect to magnetic field. 
This was a natural assumption to start with. Later,  Spitzer (1978) 
showed that additional alignment of $\bf J$ in respect to the axis of 
major inertia
should be present due to paramagnetic relaxation in the external magnetic 
field. This effect can be called Spitzer relaxation to distinguish it from
the Barnett relaxation discovered by Purcell (1979).

Here we assume that the ratio of the gas damping time 
to the time of internal relaxation is much greater than unity (see estimates
in Roberge, DeGraff \& Flaherty 1993). In this case, the
deviations of ${\bf J}$ from the axis of major inertia, $\bf a$ 
can be described thermodynamically (LR97). 

For a grain with fixed angular momentum $J\equiv |{\bf J}|$ and with
components of the moment of inertia 
related as $I_{\rm x}<I_{\rm y}<I_{\rm z}$, the kinetic energy is 
\begin{equation}
E=\frac{J^2}{2}\left [ (1/I_{\rm x}-1/I_{\rm y})\sin^2\xi \; \sin^2\theta+
(1/I_{\rm y}-1/I_{\rm z})\sin^2\theta+1/I_{\rm z}\right] ~~~ ,
\label{ii4}
\end{equation} 
where $\xi$ is the azimuthal angle in the $x-y$ plane, and $\theta$ is
the angle between $\bf J$ and $z$-axis. 
Although the equilibrium position of the grain with fixed $J^2$ corresponds
to $\theta\equiv 0$,  thermal fluctuations cause deviations from 
it. These deviations 
can be described by the Boltzmann factor $\exp\{-E/(kT_s)\}$ (LR97).

A complex precession of $\bf J$ due to these fluctuations modifies
both the grain interaction with the gaseous flow and the
dichroic absorption. However, it seems
reasonable to assume that the behaviour of such an asymmetric grain can be
approximated by the behaviour of a spheroidal grain with the moment of inertia 
somewhere between $I_x$ and $I_y$ (further on we denote this mean
moment by $I_{\bot}$). Here we also adopt this approximation 
and hope to compare the results for spheroidal and irregular grains elsewhere.

For a spheroidal grain with termprature $T_{\rm s}$, 
Eq.~(\ref{ii4}) can be simplified, and the
distribution function of the angular momentum in the grain reference frame 
is (LR97)
\begin{equation}
f_{\rm TE}(\theta_2)={\rm const}\times \sin\theta_2
\exp\left[-\frac{J^2}{2I_z kT_{\rm s}}\left\{(h-1)\sin^2\theta_2+1\right\}\right] 
~~~ ,
\label{ii5}
\end{equation}
where $\theta_2$ is the angle between ${\bf J}$ and the rotational 
symmetry axis of the spheroid, $h=I_z/I_{\bot}$ and the value of $J$ 
is different for different grains of the ensemble. 
The fact that the probability of a particular angular momentum depends 
on angle $\theta_2$  complicates the study. However, computations
in LR97 show that with a sufficient degree of accuracy
it is possible to substitute the rms value of $J$ for a
 Maxwellian angular momentum distribution in Eq.~(\ref{ii5})
to get an approximate measure of alignment for an ensemble of grains
with given rotational temperature. Therefore as the zero approximation we use
\begin{equation}
J^2=k T_{\rm eff}I_z(2/h+1) ~~~ ,
\label{ii6}
\end{equation}
(Landau \& Lifshitz 1980), where
\begin{equation}
T_{\rm eff}\approx \frac{T_{\rm g}+T_{\rm s}}{2}+\frac{1}{3}\,\frac{mu^2}{2k}~~~,
\label{ii7}
\end{equation} 
$m$ and $u$ are the mass and velocity of bombarding atoms, 
respectively, and $T_{\rm g}$ is the gas  temperature. 
We expect that numerical simulations by Roberge will
determine the accuracy of the approximation adopted.

 For hypersonic drifts in diffuse medium
discussed in Paper I, $T_{\rm eff}\gg T_{\rm s}$ and the alignment of 
$\bf J$ in the grain reference frame is close to being perfect. Indeed, if the
drift velocity of grains exceeds the velocity of atoms in diffuse 
clouds 3 times, $T_{\rm s}/T_{\rm eff}\approx 0.1$. The corresponding 
distribution  of $\bf J$ (see Eq.~(\ref{ii5})) has a peak near the axis of 
major inertia (see Fig.~1) and can be approximated by a delta-function.
This, however, may not be true for a different grain environment.
For instance in molecular clouds, gas and grains 
have similar temperatures and, for mildly supersonic drift velocities, 
$T_{\rm eff}$ is of the same order as $T_{\rm s}$.

In short, perfect alignment between $\bf J$ and $\bf a$
does not seem to be universally applicable to interstellar grains. 
Below we describe the Gold alignment when this constraint is lifted.

\section{Gold alignment for $\bf J$ not parallel to $\bf a$}

\subsection{Analytics for the Rayleigh reduction factor in the absence of
internal relaxation}

The distribution of angular momentum can be characterized by  
function $f(n,{\bf J})$, where $n$ is the number of grain-atomic collisions. 
In general, the direction of $\bf J$ should be defined by angles $\theta_{1}$
 and $\varphi_{1}$ in the ``gas reference frame'' and by $\theta_{2}$
 and $\varphi_{2}$ in the ``grain reference frame'' (see Fig.~2).
Angle  $\varphi_{1}$  describes the precession of $\bf J$ about 
magnetic field and angle $\varphi_{2}$ describes the precession of 
the spheroid's axis of rotational 
symmetry about $\bf J$. Henceforth 
grains will be approximated  by spheroids with semiaxes $a$ and $b$.

As the change of grain angular momentum in the course of an individual 
collision is small,  the alignment of $\bf J$ can be described 
by the Fokker-Planck equation 
(see Reichl 1980, Roberge, DeGraff \& Flaherty 1993). Then following  
Dolginov \& Mytrophanov (1976), we can write
\begin{equation}
\frac{\partial f({\bf  x},n)}{\partial n}=
\sum_{i=0}^{2}a_{i}({\bf  x})
\frac{\partial f({\bf  x},n)}{\partial x_{i}}+
\sum_{k,i=0}^{2}b_{ik}({\bf  x})
\frac{\partial^{2} f({\bf  x},n)}{\partial x_{i}\partial x_{k}} ~~~ ,
\label{e68}
\end{equation}
where $\bf x$ is a vector in the phase space with coordinates $ J$, 
$\cos \theta_{1}$, $\cos\theta_{2}$, $\varphi_{1}$ and $\varphi_{2}$,   
and the coefficients $a_{i}$
 and $b_{ik}$ obtained on the  assumption of supersonic grain drift 
can be found in Appendix A. The solution of Eq.~(\ref{e68}) 
in the limit of hypersonic drift is (Dolginov \& Mytrophanov 1976)
\begin{equation}
f(J,\cos\theta_{1},\cos\theta_{2},n)=
 \frac{{\rm const}_{3}}{n^{3/2}}\exp\left(
-\frac{J^{2}(1+g\cos^{2}\theta_{2}+s\cos^{2}\theta_{1})}
      {2nb^{2}p^{2}(1+g+s)}\right) ~~~ ,
\label{2.14}
\end{equation}
where
\begin{equation}
s=-\frac{1}{2}
\frac{\langle p^{2}\rangle-3\langle p_{z}^{2}\rangle}
     {\langle p^{2}\rangle-\langle p_{z}^{2}\rangle}
\label{2.10}
\end{equation}
 is the external flux anisotropy and 
\begin{equation}
g=\frac{a^{2}-b^{2}}{2b^{2}}
\label{g}
\end{equation}
is the grain non-sphericity. 
Note, that $\langle p^{2}\rangle$ and $\langle p_{z}^{2} 
\rangle$ are the averaged squared momentum and its ${\rm Z}_{1}$ 
component (see Fig.~2) transferred to a grain in an individual collision.
Both $g$ and $s$ can vary from $-0.5$ to $\infty$. It is easy to 
see that $g=-0.5$ corresponds to flakes and $g\rightarrow \infty$ to needles;
$s=-0.5$ corresponds to a flux perpendicular to magnetic field, while
$s\rightarrow \infty$ to a flux parallel to the field.
The fluxes  are measured in the grain reference frame, and therefore a 
gaseous flux with the velocity $u$ is equivalent to grain drift 
with the velocity $-u$ in respect to the ambient gas.

Spherical grains ($g=0$) correspond to $a=b$, while isotropic fluxes 
($s=0$) to $\langle p^{2}\rangle=3~\langle p_{z}^{2} \rangle$. 
We expect changes in grain alignment when $s=0$ and/or $g=0$. 
 To have a picture which is easy to visualize, we refer to fluxes (drifts)
corresponding to $s<0$ as ``fluxes (drifts) at large angles to magnetic 
field'', and to those  corresponding to $s>0$ as ``fluxes (drifts)  
at small angles to magnetic field''. Note, that ``small'' angles lie within 
the interval $[0, \arccos (1/\sqrt{3}) ]$, while ``large'' angles in 
 $[\arccos (1/\sqrt{3}), \pi/2]$, if we limit our discussion to the
first quadrant. Obviously, oblate and prolate grains correspond to
$g<0$ and $g>0$, respectively.

Calculations in Dolginov \& Mytrophanov (1976) show that the solution 
given by Eq.~(\ref{2.14}) is accurate up to $0.25|gs|$ order terms for 
$|s|<1$ and $|g|<1$. The accuracy becomes of the order of $s^{-2}$ when 
$s\rightarrow \infty$ and $|g|<0$, whereas it is 
$g^{-2}$ when $g\rightarrow \infty$ and $s<0$. If both $g$ and $s$ are 
large, the accuracy is of the order of max$[g^{-1},~s^{-1}]$.

Angle $\hat{\theta}$ can be found from simple geometric 
considerations (see Fig.~2)
\begin{equation}
\cos\hat{\theta}=\cos\theta_{1}\cos\theta_{2}+\sin\theta_{1}\sin\theta_{2}
\cos(\varphi_{1}-\varphi_{2}) ~~~ .
\label{I2.3}
\end{equation}
Averaging out the dependencies on $\varphi_{1}$ and $\varphi_{2}$
provides (see Davis \& Greenstein 1951 eq.~108):
\begin{equation}
(\cos^{2}\hat{\theta})_{\varphi}=
0.5(1-\cos^{2}\theta_{1}-\cos^{2}\theta_{2}+
3\cos^{2}\theta_{1}\cos^{2}\theta_{2}) ~~~ .
\label{2.43}
\end{equation}
To calculate the Rayleigh reduction factor given by Eq.~(\ref{2.45}), 
it is necessary to find $\langle \cos^{2}\hat{\theta}\rangle$. After 
averaging Eq.~(\ref{2.14}) over $J$ the required 
distribution function can be found to be 
\begin{equation}
W_{\rm G}(\theta_{1},\theta_{2})=
C(g,s)(1+g\cos^{2}\theta_{2}+s\cos^{2}\theta_{1})^{-\frac{3}{2}} ~~~ ,
\label{2.44}
\end{equation}
where $C(g,s)$ is the normalization constant for any   
fixed $s$ and $g$. Then 
\begin{equation}
\langle\cos^{2}\hat{\theta}\rangle=
\frac{\int \!\!\int \cos^{2}\hat{\theta}~W_{\rm G}~
\sin\theta_1\sin\theta_2 {\rm d}\theta_1 {\rm d}\theta_2}{\int \!\!\int 
W_{\rm G}~
\sin\theta_1\sin\theta_2 {\rm d}\theta_1 {\rm d}\theta_2} ~~~ .
\label{2.444}
\end{equation}
The necessary calculations are performed
in Appendix B, and here we present the final expression 
\begin{equation}
\langle\cos^{2}\hat{\theta}\rangle=\frac{1}{2gs}
\left(1+g+s+gs-C(g,s)\sqrt{1+g+s}\right) ~~~ ,
\end{equation}
where
\begin{eqnarray}
C(g,s)=\left\{
\begin{array}{ll}
\,\sqrt{gs}\,\,\left(\arctan\sqrt{\frac{gs}{1+g+s}}\right)^{-1},&
gs>0~~~ ,\\
\\
\sqrt{|gs|}\left({\rm arctanh}\sqrt{\frac{|gs|}{1+g+s}}\right)^{-1},&
gs<0~~~ ,\\
\end{array}\right. 
\label{cgs}
\end{eqnarray}
which is valid for all possible values of $s$ and $g$. This expression can
be compared with the analytics found for the case of strong relaxation (see
Appendix C). 

The corresponding Rayleigh reduction factor is 
\begin{equation}
R=\frac{1}{4}+\frac{3\sqrt{1+g+s}}{4gs}\left(\sqrt{1+g+s}-C(g,s)\right) 
~~~ .
\label{3.18}
\end{equation}
This measure
enters the formulae for intensity of polarized radiation due to dichroic 
absorption. Equation~(\ref{3.18}) encompasses a variety of circumstances,
which are explored below.

\subsection{Comparison with Paper I}

 Here we discuss the alignment for various values of $s$ and $g$. When 
internal dissipation aligns $\bf J$ and the axis of major inertia perfectly, 
Eqs.~(\ref{2.43}), (\ref{2.45}) and (\ref{j1}) give the following 
relation between the Rayleigh reduction factor and the measure of alignment 
of angular momentum, $\sigma_{J}$, used in Paper~I
\begin{equation}
R=\left\{\begin{array}{cl}
\sigma_{J}, & {\rm for~oblate~grains} ~~~ ,\\
-0.5\sigma_{J}, & {\rm for~prolate~grains} ~~~ .\\
\end{array}\right. 
\label{3.22}
\end{equation}
It is easy to see that prolate and oblate grains produce polarization of 
the same sign despite the fact, that the Rayleigh
reduction factor has opposite signs in these two cases.
This becomes clear if one recalls that oblate grains aligne
their short axes 
in respect to the magnetic field, while prolate grains 
align their long axes.

First, consider oblate grains ($ g<0$) subjected to drift at large 
angles to magnetic field ($s<0$). The corresponding 
measure of alignment is shown in Fig.~3. A comparison with 
fig.~3 in Paper I reveals the decrease of alignment associated with 
suppression of internal dissipation. For instance, for the drift 
perpendicular to magnetic field lines, this measure for flake-like 
grains is 0.25 if the Barnett alignment is absent, while it 
reaches unity (complete alignment) if the internal dissipation is
efficient (see Fig.~4).

The alignment measure for oblate grains drifting at small angles to 
magnetic field ($s>0$) is shown in 
Fig.~6. The comparison with fig.~4  in Paper I testifies an 
order of magnitude decrease of the measure. A cross-section of the plot 
for $g=-0.5$ (see Fig.~5.) shows that  flakes are only marginally 
aligned. This is in contrast with the case of intense internal 
dissipation, when flakes are well aligned and the measure of 
alignment approaches $-0.5$, if the flux and magnetic field directions 
coincide (Lazarian 1994a). Due to the symmetry of the distribution function 
given by Eq.~(\ref{2.14}) under 
a simultaneous interchange $\cos\theta_{1} \leftrightarrow \cos\theta_{2}$ and 
$s \leftrightarrow g$, we can claim that needles are
marginally aligned if grain drift is perpendicular to the field,
i.e. ${\bf u}\bot {\bf B}$. Indeed, the corresponding measure of alignment 
for prolate grains is negligible (see Fig.~7). In breif, if
the internal dissipation is negligible, prolate and oblate 
grains are marginally aligned for drifts at angles close to $\pi/2$ and 0, 
respectively.

When both $s$ and $g$ are positive, the corresponding measure of 
alignment is shown in Fig.~8. One can see that this measure 
tends to $+0.25$ when $s$ or/and $g$ tend to infinity. In terms of 
$\sigma_{J}$ (see Eq.~(\ref{3.22})) this is equivalent to $\sigma_{J}=-0.5$ 
obtained in Paper~I. This correspondence has a simple explanation. Indeed,
 for a needle the angular momentum should be 
directed along the axis of major inertia even in the absence of internal 
dissipation.

The requirement of large $s$ and $g$ places stringent 
constrains on the efficiency of alignment for typical ISM conditions.
Indeed, let grain drift velocity components perpendicular and parallel 
to magnetic field be  $u_{\bot}$ and $u_{\|}$, respectively, then if 
$u_{\bot}$ is much greater than the rms velocity of gas atoms $v_{rms}$ 
\begin{equation}
s=\frac{2-w^{2}}{2~w^{2}} ~~~ ,
\label{perp1}
\end{equation} 
where $w\approx u_{\bot}/u_{\|}$. If $u_{\bot}\ll~v_{rms}$, 
than
$w\approx v_{rms}/v_{\|}$ should be used in Eq.~(\ref{perp1}). 
In any case, $w$ is unlikely to be less than $0.1$, and therefore $s$ is not
likely to be greater than 100. We neither believe that the axis ratio of 
a typical prolate grain exceeds 10, therefore $g$ is likely to be 
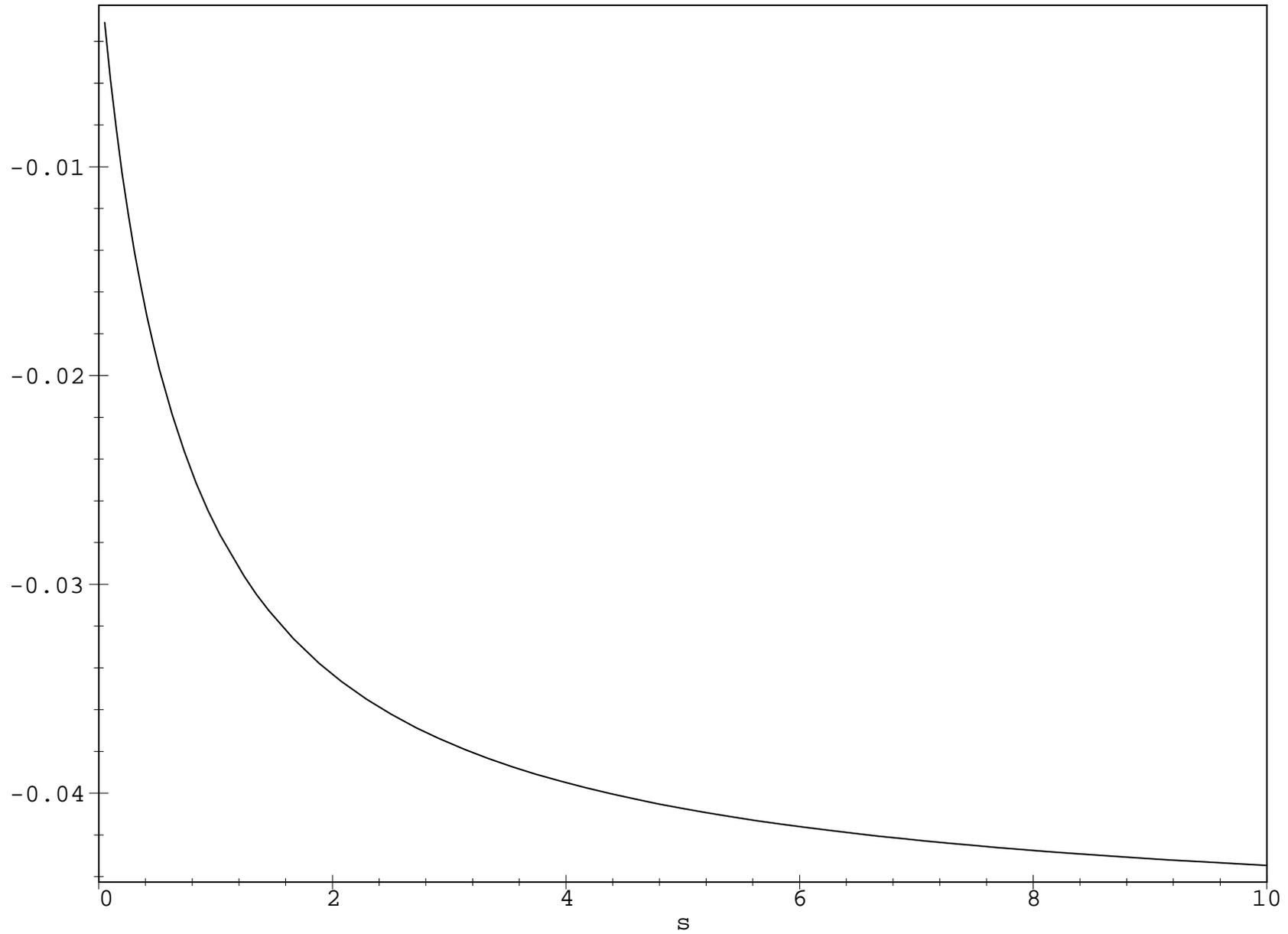
less than 50. The measure of alignment for $w\in [0.2, 0.7]$  and 
the axis ratio $y\equiv a/b\in [1,10]$ is shown in Fig.~9. 
The cross-sections of the plot  for $y=5$ and $y=10$ are shown in Fig.~10.

To summarize, in the absence of  internal dissipation, the measure of 
 alignment drops. Oblate 
grains are most aligned for ${\bf u}\bot {\bf B}$, and prolate grains
are most aligned for ${\bf u}\| {\bf B}$.

\section{Generalized problem}

Up to now two extreme cases were discussed: strong internal dissipation,
when the angular momentum is coupled to the  axis of major inertia, 
and weak internal dissipation, when  the residual alignment of
angular momentum is due to the differences between the maximal and minimal
 rotational inertia. In both cases analytical solutions were
found. These solutions provide the 
upper and lower bounds for the measure of  alignment. For instance, 
Fig.~11 shows these two bounds for ${\bf u}\bot {\bf B}$ when 
the alignment is caused by ambipolar diffusion. A conspicuous feature 
of this particular figure is that the two plots are different
for ideal spheres, i.e. when no axis alignment is expected.
Equation~(\ref{ii5}) testifies that for $h\rightarrow 1$, all positions of
$\bf J$ become equally probable. This  fact was
ignored in our  simplified approach adopted  in Paper~I. 
Fig.~11 also testifies,
 that the internal dissipation strongly influences grain
alignment, as the spread in Rayleigh reduction factors for the two extreme
cases is wide.

To account for the incomplete alignment of $\bf J$ in the 
grain reference frame, 
one has to incorporate internal dissipation in the Fokker-Planck
equation. An analytical study  in this case seems formidable, 
and a numerical approach, e.g. similar to the one used by Roberge, 
DeGraff \& Floherty (1993) may be advantageous. For such a study both 
the analytical solutions obtained above and those derive 
in Paper I should serve as benchmarks.

 A less rigorous, but less laborious way to account for the incomplete 
alignment  is to follow Jones \& Spitzer (1967).
Let the distribution function of $\bf J$
in the presence of internal dissipation, $W_{\rm GD}$, be the product of the 
distribution functions $W_{\rm G}$ and $W_{\rm D}$ given by Eqs~(\ref{2.44}) 
and (\ref{ii5}), respectively. Then after 
expressing $h$ in terms of $g$ and $J^2$ in terms of $T_{\rm eff}$ using 
Eqs~(\ref{ii6}) and (\ref{ii7}), respectively, we get
\begin{equation}
W_{\rm GD}\approx {\rm const}\times \sin\theta_2~
(1+g\cos^{2}\theta_{2}+s\cos^{2}\theta_{1})^{-\frac{3}{2}}\exp\left (
\frac{T_{\rm eff}}{T_{\rm s}} \frac{g(2g+3)}{g+1} \frac{\sin^2\theta_2}{2}\right ) 
~~~ .
\end{equation}

Using $W_{\rm GD}$ in Eq.~(\ref{2.444}) 
we calculated  $\langle\cos^{2}\hat{\theta}\rangle$  and the Rayleigh
reduction factor for  $T_{\rm eff}/T_{\rm s}$ equal to
100, 10 and 1.1. For example, Fig.~12  shows the Rayleigh reduction 
factor as a function of grain oblateness. The comparison between Figs~12 and
11 shows that while for efficient internal relaxation corresponding
to $T_{\rm eff}/T_{\rm s}>100$ the approximation of perfect coupling is 
appropriate, in the case of $T_{\rm rot}\approx T_{\rm s}$, this is no longer
true, and the analytics disregarding internal relaxation provide a better
fit. Similar conclusions are valid for other values of $s$ and $g$
with the exception of $g\rightarrow 0$. In this case, as 
 discussed earlier, there is no coupling between the angular momentum 
and the axis of greatest inertia.

The alignment of prolate
grains by ambipolar diffusion is marginal and the comparison between Figs~13
and 7 shows that the internal dissipation does not change it much. 
In contrast, Figs~14 and 5 show that the internal
dissipation drastically changes the alignment of oblate grains due
to the radiation pressure. If this alignment is marginal when
the internal relaxation is suppressed (see Fig.~5), it becomes substantial
as soon as the internal relaxation is present. This difference in 
the susceptibility to internal dissipation is 
easy to understand if one recalls that without internal dissipation
$\bf J$ is only marginally aligned with the axis of major inertia
for oblate grains and the alignment is substantial in the 
case of prolate grains. Therefore Fig.~14 shows a marginal difference 
in grain alignment for different values of internal dissipation as grains
become sufficiently prolate.  

To summarize, our results  show that Gold alignment is modified by internal 
relaxation, and to predict the alignment accurately it is essential
to estimate the ratio of the effective rotational grain temperature to
grain material temperature. Small temperature ratios  usually occur
when the grain drift and the thermal velocities are 
comparable.\footnote{We disregard a rather artificial case of  hot 
grains drifting in cold gas.} Then it is essential to account for the fact, 
that,
in the grain reference frame, atoms move at the drift velocity, which 
is modified by thermal motion. In other words, grain - gas collisions
formally correspond to a range of values of $s$. To account for
this effect, we suggest to integrate over the corresponding range of 
$s$ (see Lazarian 1995a). Since there is no direct numerical calculations 
of the measure of alignment for the case of incomplete internal relaxation, 
in what follows we compare our results only with numerics 
for perfect coupling between $\bf J$ and the axis of major inertia.

\section{Comparison with RHM}

A comprehensive numerical study of the Gold alignment for perfect Barnett
relaxation was done in RHM. The authors presented a 
detailed study of grain alignment for a range of drift velocities 
starting from subsonic ones.  Paramagnetic relaxation  was also
included in their model. Here we briefly
discuss how to improve our model to include  both subsonic
drift and paramagnetic relaxation. 

\subsection{Subsonic drift}

The model adopted here assumes that the grain drift is essentially
hypersonic. In other words, we have ignored the rms velocity
of gaseous atoms  as compared with the drift velocity and assumed that
atoms hit the grain from one direction defined by 
\begin{equation}
\phi=\arcsin\sqrt{\frac{u_{\rm x}^2+u_{\rm y}^2}
                       {u_{\rm x}^2+u_{\rm y}^2+u_{\rm z}^2}} ~~~ ,
\label{cc1}
\end{equation}
where $u_{\rm i}$, i$={\rm x,y,z}$, are the components of atom drift 
velocity in the grain reference frame. 
Obviously, the anisotropy parameter $s$ can be expressed as a function 
of  $\phi$
\begin{equation}
s=\frac{~~~1-3\cos^2\phi}{1-\cos^2\phi}
\label{cc3}
\end{equation}
and hence the Rayleigh reduction too.

When the drift is subsonic, in the grain reference frame atoms are viewed
as approaching from various directions, any particular atom at angle
\begin{equation}
\phi=\arcsin\sqrt{\frac{(u_{\rm x}+v_{\rm x})^2+(u_{\rm y}+v_{\rm y})^2}
                       {(u_{\rm x}+v_{\rm x})^2+(u_{\rm y}+v_{\rm y})^2
+(u_{\rm z}+v_{\rm z})^2}}~~~,
\label{cc2}
\end{equation}
 where $v_{\rm i}$,  i$={\rm x,y,z}$, are the components of rms velocity of the 
atom. According to Eq.~(\ref{cc2}), angle $\phi$  varies from atom to atom 
due to variations in $v_{\rm i}$. To obtain the statistics of
$R$ we subdivide atoms into subgroups of atoms having the same 
$v_{\rm i}$.\footnote{As the time of alignment is 
much shorter than the damping time (see Paper~I), each of the subgroups may
by itself cause alignment.} For each of the subgroups, 
$\phi$ is constant, hence the measure of alignment can be obtained by
averaging $R(u_{\rm i},v_{\rm i})$ over a Maxwell-Boltzmann distribution 
of $v_{\rm i}$.

Elsewhere we hope to test the applicability of such an approach for a 
considerable range of drift velocities by comparing our predictions with direct
numerical simulations. At the moment, we want to show that our estimates
 are in a reasonable agreement with the data presented in RHM. 

RHM assume that the angular momentum is perfectly coupled with the
axis of major inertia, which corresponds to $T_{\rm s}/T_{\rm eff}=0$ in 
our model. The effect of the spread
of atom velocities in the grain reference frame should not
depend upon the position of the angular momentum in respect to the axis of 
the major inertia. Indeed, the velocities of particles on the grain surface
are much smaller than the velocities of striking atoms and therefore
if we obtain the correspondence between our and RHM predictions we may hope 
that our treatment is applicable when the internal alignment is incomplete.

Figure~8 in RHM shows the Rayleigh reduction factor for oblate grains
drifting with different velocities in respect to gas. The calculations
are made up to Mach number 4, but the saturation of the alignment is 
obvious from this plot. Therefore we will consider that these values
correspond to hypersonic velocities and define
 $R(\phi_0)$, where $\phi_0$ in the case of ambipolar 
diffusion studied in RHM is $\pi/2$.
For our simplified estimates we observe that for large Mach numbers $M$, 
Eq.~(\ref{cc2}) provides
\begin{equation}
\phi\approx \arcsin\left(1-0.5~M^{-2}\right)\approx 
\frac{\pi}{2}-\frac{1}{2}M^{-2}
-\frac{\sqrt{2}}{24}M^{-3} ~~~ ,
\label{cc4}
\end{equation}
where the first term corresponds to $\phi_0$ and the rest can be interpreted
as $\delta\phi$. Therefore
\begin{equation}
\langle R\rangle_{\phi}\approx \frac{1}{2\delta\phi}
\int^{\phi_0+\delta\phi}_{\phi_0-\delta\phi}R(\phi){\rm d}\phi \approx
R(\phi_0)+2\frac{{\rm d}R}{{\rm d}s}\frac{{\rm d}s}{{\rm d} \phi}
\delta\phi ~~~ ,
\label{cc5}
\end{equation}
where 
\begin{equation} 
R(\phi_0)=-3.5 - 3\,g + 
3\,\sqrt {1+2\,g}\,\,(1+g)\,\arcsin{\frac {1}{2\,\sqrt {1+g}}}~~~,
\label{cc6}
\end{equation}
\begin{eqnarray}
\left.\frac{{\rm d}R}{{\rm d}s}\right|_{s=-0.5}
&=&
 \frac{3+3\,g}{\sqrt {1+2\,g}\sqrt {3+4\,g}}
 \left[
-1  - 2\,g - 2\,\sqrt {1+2\,g}\sqrt {3+4\,g}
\right.\nonumber\\
& & \nonumber\\
& & \left. + ~\sqrt {3+4\,g}\,(4+6\,g)\,
\arcsin\left({\frac {1}{2\,\sqrt {1+g}}}\right)
\right]~~~,
\label{cc7}
\end{eqnarray}
and 
\begin{equation}
\frac{{\rm d}s}{{\rm d}x}=
\frac {6\,\cos x\,\sin x}{1-\cos^2 x}-
\frac {\left (2-6\,\cos^2 x\right )\cos x\,\sin x}
       {\left (1-\cos^2 x\,\right )^2}~~~,
\label{cc8}
\end{equation}
with ${\rm d}s/{\rm d}x \approx 0.032$ when $x=\phi=\pi/2$.

For instance, RHM find that at large Mach numbers the Rayleigh reduction 
factor for grains with the axis ratio 0.5 is $\approx~0.25$. This axis ratio 
 corresponds to $g=1.7$, when pluged into Eq.~(\ref{cc5}) it gives 
$\langle R\rangle_{\phi}\approx 0.21$ for $M=2$. This is comparable with
the value $\approx 0.19$ that follows from fig.~8 in RHM. Similarly
for the axis ratio $0.25$ corresponding to $g=7.5$, 
$\langle R\rangle_{\phi}\approx 0.27$, for $M=2$, 
which is of the same order, as the result in RHM 
($\approx 0.26$).\footnote{Again we take the values of the Rayleigh 
reduction factor
obtained for $M=4$ in RHM and substitute in our formulae. For the axis
ratio $0.25$ this value $\approx 0.39$.}
As the values of the Rayleigh reduction
factor obtained in RHM for high Mach numbers essentially correspond to
the values obtained in the analytical treatment in Paper~I (see fig.~12 in RHM)
it is possible to see, that a purely analytical treatment is appropriate
at least for some values of subsonic drift velocities. The entire range of
velocities will be treated elsewhere. 

Our estimates above were obtained using the analytical solutions
obtained in Paper~I for perfect coupling of $\bf J$ with the axis of 
major inertia. Evidently our approach is also applicable to describe
alignment at low Mach numbers using the analytics
and ``semi-analytics'' obtained in Sections 3 and 4 respectively.
We are looking forward to the progress in numerical techniques to be able
to compare our predictions with the results of  direct numerical 
simulations.

\subsection{Mechanical and paramagnetic alignment}

So far we have talked only about mechanical alignment and 
completely disregarded the paramagnetic one. This is justifiable only if 
 the Davis-Greenstein alignment is negligible. In general,
 the Davis-Greenstein alignment must be accounted for.
To do this we propose a simple formular.

If we denote the Rayleigh reduction factor of the Gold alignment by 
$R_{G}$, the measure of the internal alignment by $Q_X$, and
that of Davis-Greenstein process by $R_{DG}$,
it is possible to estimate the measure of the overall alignment as 
\begin{equation}
R^{\Sigma}\approx Q_X\frac{Q_X\sigma_{G}+R_{G}R_{DG}+
Q_X R_{DG}}{Q_X^{2}+2R_{G}R_{DG}} ~~~.
\label{cc9}
\end{equation}
To obtain the expression above we used the expression for $J$ alignment
given by  eq.~45 in Lazarian (1995a) and the approximation
\begin{equation} 
R_i\approx Q_{J(i)}\times Q_X~~~,
\end{equation}
where $R_i$ is the Rayleigh reduction factor obtained for the $i$
mechanism acting alone, while $Q_{J(i)}$ is the measure of $J$ alignment
relatively to magnetic field. The latter approximation follows from spherical 
trigonometry (see Eq.~(\ref{2.43})). Indeed, if two processes are 
independent, then 
 \begin{equation}
\langle \cos^{2}\theta_1 \cos^{2}\theta_2\rangle \approx 
\langle \cos^{2}\theta_1\rangle
\langle \cos^{2}\theta_2\rangle~~~.
\end{equation}
The approximation above was proved to be sufficiently accurate for the
Davis-Greenstein process in Lazarian (1995a), but may be much less accurate
for the Gold alignment. In any case, we treat Eq.~(\ref{cc9}) only as 
a conjecture to be tested in future.

A study of the simultaneous action of the paramagnetic and mechanical
alignment has been done recently in RHM for $T_d/T_g=0$.
In this case, $\sigma_{B}\equiv 1$ and Eq.~(\ref{cc9}) reduces 
to the one derived in Lazarian (1995a).

First of all, to provide the comparison we must define the measure of
alignment for the Davis-Greenstein process when grains are 
subjected to a supersonic flow. Naturally,
as grains drift faster the rate at which atoms arrive to their surface
increases, and therefore the gaseous damping time decreases. 
For our rough estimate of the Davis-Greenstein alignment subjected
to the supersonic flow, we simply substitute overall velocity instead of the
thermal one into the expressions for the diffusion coefficients for gaseous 
bombardment (see Roberge, DeGraff \& Flaherty 1993). This changes both 
 $T_{\rm eff}$ and the ratio of the
gaseous and magnetic damping times $\delta_1$ (see eq.~25 in Lazarian 1995).
The first change is not relevant here because to compare our estimates with 
the calculations in RHM, we assume $T_{\rm s}/T_{\rm eff}=0$. On the contrary, 
the second change rescales $\delta_1$ calculated for the non-drifting
grain (this value is shown in figs~4 and 9 in RHM). Let the rescaled 
value be $\delta/M$. To obtain $\sigma_G$, we use the zero approximation in 
Lazarian (1995c).  

	For spherical grains when $\delta_1=10$, RHM obtain $\sigma\approx
0.40$ (see fig.~4 in RHM). Our approximation gives $\sigma\approx 0.38$. 
It is easy to see from fig.~9 in RHM that for $\delta_1=1$, 
axis ratio 0.5 gives $\sigma\approx 0.32$, whereas axis ratio $0.25$ gives 
 $\sigma\approx 0.48$. Our approach provides $\sigma\approx 0.29$ and 0.46, 
respectively. This approximate correspondence let us hope that our simplified
analytical treatment reproduces essential features of the alignment.

\section{Discussion}

In short, we have shown  that the alignment of $\bf J$ in the grain
reference frame, that arises from the difference between the 
 grain material and rotational temperatures, is essential for 
the Gold alignment.
When velocities of grain drift are hypersonic, we can assume perfect
coupling of $\bf J$ and the axis of major inertia. However, this assumption 
fails for mildly supersonic drifts when the gas and grain
temperatures are comparable. Such conditions are expected, e.g. in molecular
clouds undergoing ambipolar diffusion. Then incomplete alignment between 
$\bf J$ and  the axis of major inertia should be accounted for. 

Our results also show that the effect of incomplete relaxation
is more vivid for oblate grains than for prolate ones. This is a 
consequence of the fact, that for sufficiently prolate grains the alignment
of $\bf J$ in respect to grain axis of major inertia is manifest even
without internal relaxation.

The analytical results obtained above and those derived in Paper~I provide 
the lower and upper bounds for Gold alignment. In a general case of 
incomplete internal alignment, we suggested a semi-analytical approach. 

To provide a quantitative description of Gold alignment for drift 
velocities comparable with the thermal velocities of gaseous atoms,
we suggested an expansion of Rayleigh reduction factor in a 
series over the drift Mach number and obtained a fair correspondence
with the results of direct numerical computations.

\acknowledgements

Comments by Bruce Draine initiated  this work, but its completion would not be 
possible if not for the encouragement by Ethan Vishniac. 
I am indebted to Wayne Roberge for his constructive criticism of 
the initial draft.  Present work was
partially done during my stay in the stimulating environment of
Harvard-Smithsonian Observatory and I gratefully acknowledge a Short-Term
Visitor Fellowship from the Smithsonian Institution and hospitality of
Phil Myers.  The research is
supported by  NASA grant NAG5 2773 at the University of Texas at Austin.

\appendix

\section{Coefficients of the Fokker-Planck equation \label{A.1}}

The coefficients $a_{i}$ and $b_{ik}$ of the Fokker-Planck equation 
are determined by the change of the grain angular momentum  due to 
grain - atom  collision
\begin{eqnarray}
a_{0}&=&\left\langle
\frac{1}{2}\bigtriangleup x_{2}
\frac{\partial \bigtriangleup x_{1}}
     {\partial  x_{2}}+
\frac{1}{2}\bigtriangleup \varphi_{1}
\frac{\partial \bigtriangleup x_{1}}
     {\partial  \varphi_{1}}-
\bigtriangleup x_{1}\right\rangle ~~~, \label{10}\\
a_{m}&=&\left\langle
\frac{1}{2}\bigtriangleup x_{1}
\frac{\partial \bigtriangleup x_{m}} 
     {\partial  x_{1}}+
\frac{1}{2}\bigtriangleup x_{m}
\frac{\partial \bigtriangleup x_{m}}
     {\partial x_{m}}\right.\nonumber\\
     & & +\, \left.
\frac{1}{2}\bigtriangleup \varphi_{m}
\frac{\partial \bigtriangleup x_{m}}
     {\partial \bigtriangleup \varphi_{m}}-
\bigtriangleup x_{m}\right\rangle ~~~, \label{11}\\
b_{ik}&=&\left\langle
\bigtriangleup x_{i}\bigtriangleup x_{k}\right\rangle ~~~, \label{12}
\end{eqnarray}
where $m=1,2$, $\;\;k,i=0,1,2$, and the angular brackets denote averaging over
atom impacts over the grain surface. The quantities $\bigtriangleup 
x_{i}$ and $\bigtriangleup \varphi_{i}$  and the corresponding 
$a_{i}$ and $b_{ik}$ were calculated in Dolginov \& Mytrophanov 
(1976) using the following equations: 
\begin{eqnarray}
\bigtriangleup x_{j}&=&\bigtriangleup({\bf  e}_{j}\cdot{\bf  J})-
({\bf  e}_{j}\cdot{\bf  J})\bigtriangleup J ~~~ , \nonumber\\
\bigtriangleup x_{0}&=&{\bf  j}\cdot\bigtriangleup{\bf  J} ~~~ , \nonumber\\
\bigtriangleup \varphi_{j}&=&(({\bf  e}_{j}\times{\bf  j})\cdot \bigtriangleup
{\bf  J}(J(1-x_{j}^{2}))^{-1} ~~~, \nonumber
\end{eqnarray}
where $j=1,2$, ${\bf  e}_{1}={\bf  H}/|{\bf  H}|$ is a unit vector along 
the magnetic field, ${\bf  e}_{2}={\bf  a}/|{\bf  a}|$ is a unit 
vector along ${\rm Z}_{1}$-axis of the grain, ${\bf  j}={\bf  J}/|{\bf J}|$ 
is a unit vector along ${\bf  J}$, and 
$\bigtriangleup {\bf  J}={\bf  r}\times{\bf  p}$ with $\bf r$ and $\bf p$
used for the vector to the point of atomic impact and atom momentum,
respectively. 

To find coefficients $a_{k}$ ($k=0,1,2$) and $b_{ik}$ ($i,k=0,1,2$) (see 
Dolginov \& Mitrophanov 1976, eq.~5) one 
has to substitute $\bigtriangleup x_{j}$, $\bigtriangleup x_{0}$, 
$\bigtriangleup \varphi_{j}$ into Eqs~(\ref{10}), (\ref{11}), and 
(\ref{12}) and perform  the necessary 
averaging. Apart from averaging over the surface area  exposed to the 
flux, one has to average over the angles of  precession of 
of ${\bf  J}$ about ${\bf  e}_{2}$ and ${\bf m}$.

\section{Computation of integrals}

Equation (\ref{2.444}) involves integration of Eq.~(\ref{2.44})
with the distribution function  $W_{\rm G}(\theta_{1},\theta_{2})$. 
As a result, we get 
\begin{equation}
\langle\cos^{2}\theta\rangle=
0.5\,\left(1-\langle\cos^{2}\theta_1\rangle-
\langle\cos^{2}\theta_2\rangle + 
3\langle\cos^{2}\theta_1\cos^{2}\theta_2\rangle
\right)~~~,
\label{expl}
\end{equation}
where the explicit expression for the first term is 
\begin{equation}
\langle\cos^{2}\theta_{1}\rangle=C(g,s)
\int_{0}^{\pi/2}{\rm d}\theta_{1}\int_{0}^{\pi/2}
\frac{\cos^{2}\theta_{1}\sin\theta_{1}\sin\theta_{2}}
     {(1+s\cos^{2}\theta_{1}+g\cos^{2}\theta_{2})^{3/2}}~{\rm d}\theta_{2}
 ~~~. 
\label{sh4}
\end{equation}
After integrating over $\theta_2$ and the change of variables, 
$\sin\theta_1=x$, we have 
\begin{equation}
\langle\cos^{2}\theta_{1}\rangle=C(g,s)\int_{0}^{1}
\frac{x^{2}{\rm d}x}{(1+sx^{2})\sqrt{1+sx^{2}+g}} ~~~,
\label{I2.8}
\end{equation}
where
\begin{equation}
C(g,s)\int_{0}^{1}
\frac{{\rm d}x}{(1+sx^{2})\sqrt{1+sx^{2}+g}}\equiv 1 ~~~. 
\end{equation}
Using the identity 
\begin{equation}
\frac{1}{s}(1+sx^{2})-\frac{1}{s}=x^{2} ~~~, 
\end{equation}
Equation (\ref{I2.8}) reduces to 
\begin{equation}
\langle\cos^{2}\theta_{1}\rangle=C(g,s)\frac{1}{s}\int_{0}^{1}
\frac{{\rm d}x}{\sqrt{1+sx^{2}+g}}-\frac{1}{s} ~~~, 
\end{equation}
which can be calculated [Gradshtein \& Ryzhik 1965, 2.271(4)] to give
\begin{eqnarray}
i_{1}=
\int_{0}^{1}\frac{{\rm d}x}{\sqrt{(1+g)+sx^{2}}}=
\left\{
\begin{array}{ll}
\frac{1}{\sqrt{s}}\ln~\frac{\sqrt{s}+\sqrt{1+g+s}}{\sqrt{1+g}},
& s>0 ~~~, \\
& \\
\frac{1}{\sqrt{-s}}~{\rm arcsin}\sqrt{-\frac{s}{1+g}},
& s<0 ~~~.
\end{array}
\right.
\end{eqnarray}
For uniform representation, one can also use  inverse 
hyperbolic function for $s<0$, namely
\begin{equation}
i_{1}=\frac{1}{\sqrt{s}}~{\rm arcsinh}\sqrt{\frac{s}{1+g}}, 
\end{equation}
where 
\begin{equation}
{\rm arcsinh}~z=\ln (z+\sqrt{z^2+1})=\frac{1}{i}
{\rm arcsin}~(iz).
\end{equation}
To find $C(g,s)=i_2^{-1}$, we calculate
\begin{equation}
i_{2}=\int_{0}^{1}\frac{{\rm d}x}{(1+sx^{2})\sqrt{1+sx^{2}+g}} ~~~. 
\label{i2}
\end{equation}
By substituting $u=x^{2}+s$ into Eq.~(\ref{i2}) and evaluating the 
resulting integral (Gradshtein \& Ryzhik 1965, 2.224(5)), we have
\begin{eqnarray}
i_{2}=\left\{
\begin{array}{ll}
\frac{1}{2\sqrt{-gs}}~\ln
\frac{\sqrt{1+s+g}+\sqrt{-gs}}{\sqrt{1+s+g}-\sqrt{-gs}},
&gs<0 ~~~ ,\\
& \\
\frac{1}{\sqrt{gs}}~{\rm arctan}\sqrt{\frac{gs}{1+s+g}},
&gs>0 ~~~. \end{array}
\right.
\end{eqnarray}
 Using the inverse hyperbolic function 
\begin{equation}
{\rm arctanh}~z=\frac{1}{2}\ln \frac{1+z}{1-z}=
{\rm arctan}~(iz) ~~~, 
\end{equation}
the expression for $i_2$ can be rewritten as 
\begin{equation}
i_{2}=\frac{1}{\sqrt{-gs}}~{\rm arctanh}\sqrt{\frac{-gs}{1+s+g}}, ~~~
sg<0 ~~~. 
\label{I2.17}
\end{equation}
Finally, we obtain,\\
 for $s<0$ and $g<0$,
\begin{equation}
\langle\cos^{2}\theta_{1}\rangle=
\frac{\sqrt{-g}~{\rm arcsin}\sqrt{-\frac{s}{1+g}}}
     {s~{\rm arctan}\sqrt{\frac{gs}{1+s+g}}}-\frac{1}{s} ~~~, 
\end{equation}
for $s<0$, and $g>0$,
\begin{equation}
\langle\cos^{2}\theta_{1}\rangle=\frac{\sqrt{g}~{\rm arcsin}
\sqrt{-\frac{s}{1+g}}}
     {s~{\rm arctanh}\sqrt{-\frac{gs}{1+s+g}}}-\frac{1}{s} ~~~, 
\end{equation}
for $s>0$, and $g<0$,
\begin{equation}
\langle\cos^{2}\theta_{1}\rangle=\frac{\sqrt{-g}~{\rm arcsinh}
\sqrt{\frac{s}{1+g}}}
     {s~{\rm arctanh}\sqrt{-\frac{gs}{1+s+g}}}-\frac{1}{s} ~~~, 
\end{equation}
for $s>0$, and $g>0$,
\begin{equation}
\langle\cos^{2}\theta_{1}\rangle=\frac{\sqrt{g}~{\rm arcsinh}
\sqrt{\frac{s}{1+g}}}
     {s~{\rm arctan}\sqrt{\frac{sg}{1+s+g}}}-\frac{1}{s} ~~~, 
\end{equation}
which covers all cases. Similarly, for the second term in 
Eq.~(\ref{expl}) we obtain
\begin{equation}
\langle\cos^{2}\theta_{2}\rangle=\left\{
\begin{array}{ll}
C(g,s)\frac{1}{g\sqrt{-g}}~{\rm arcsin}\sqrt{-\frac{g}{1+s}},
& g<0 ~~~, \\
\\
C(g,s)\frac{1}{g\sqrt{g}}~{\rm arcsinh}\sqrt{\frac{g}{1+s}},
& g>0 ~~~. \\
\end{array}\right. 
\end{equation}
The last term in Eq.~(\ref{expl}) is
\begin{equation}
\langle3\cos^{2}\theta_{1}\cos^{2}\theta_{2}\rangle=3C(g,s)
\int_0^{\pi/2}\int_0^{\pi/2}
\frac{\cos^2\theta_1\, \cos^2\theta_2\, \sin\theta_1\, \sin\theta_2\, 
      {\rm d}\theta_1\,{\rm d}\theta_2}
     {(1+s\,\cos^2\theta_1+g\,\cos^2\theta_2)^{3/2}}~~~,
\end{equation}
which after obvious substitutions takes the form 
\begin{equation}
\langle3\cos^{2}\theta_{1}\cos^{2}\theta_{2}\rangle=3C(g,s)
\int_{0}^{1}x^{2}{\rm d}x\int_{0}^{1}
\frac{y^{2}{\rm d}y}
     {(1+sx^{2}+gy^{2})^{3/2}} ~~~.
\label{sh12}
\end{equation}
Let $s<0$ and $g<0$, then 
\begin{equation}
\begin{array}{ll}
s=-b^{2} ~~~,\\
g=-d^{2} ~~~,\\
\end{array}
\end{equation}
for some non-zero $b$ and $d$.
Which gives for the inner integral in Eq.~(\ref{sh12})
\begin{eqnarray}
G_{1}
&=&\int_{0}^{1}
\frac{y^{2}{\rm d}y}
     {(1-b^{2}x^{2}-d^{2}y^{2})^{3/2}}=
\left[\frac{y}{d^{2}\sqrt{1-b^{2}x^{2}-d^{2}y^{2}}}-
\frac{1}{d^3}\,{\rm arcsin}
\frac{{\rm d}y}{\sqrt{1-b^{2}x^{2}}}\right]_{0}^{1}\nonumber\\
&=&
\frac{1}{d^{2}\sqrt{1-b^{2}x^{2}-d^{2}y^{2}}}-\frac{1}{d^{3}}
{\rm arcsin}\frac{d}{\sqrt{1-b^{2}x^{2}}}~~~.
\end{eqnarray}
Therefore
\begin{equation}
\langle3\cos^{2}\theta_{1}\cos^{2}\theta_{2}\rangle=\frac{3C(g,s)}{d^{2}}
\left\{\int_{0}^{1}\frac{x^{2}{\rm d}x}{\sqrt{1-d^{2}-b^{2}x^{2}}}-
\frac{1}{d}\int_{0}^{1}x^{2}~{\rm arcsin}\frac{d}{\sqrt{1-b^{2}x^{2}}}
{\rm d}x\right\} ~~~.
\end{equation}
The second integral in the square brackets can be integrated by parts
\begin{eqnarray}
G_{2}
&=&\frac{1}{d}\int_{0}^{1}x^{2}
~{\rm arcsin}\frac{d}{\sqrt{1-b^{2}x^{2}}}{\rm d}x\nonumber\\
&=&\frac{1}{3}\left[\frac{1}{d}~{\rm arcsin}\frac{d}{\sqrt{1-b^{2}x^{2}}}-
\frac{db^{2}}{d}
\int_{0}^{1}\frac{x^{4}{\rm d}x}
                 {(1-b^{2}x^{2})\sqrt{1-d^{2}-b^{2}x^{2}}}\right] ~~~,
\label{sh16}
\end{eqnarray}
where
\begin{equation}
\left({\rm arcsin}\frac{d}{\sqrt{1-b^{2}x^{2}}}\right)'=
\frac{db^{2}x}{(1-b^{2}x^{2})\sqrt{1-d^{2}-b^{2}x^{2}}}
\end{equation}
was taken into account. The last integral in Eq.~(\ref{sh16}) 
can be evaluated using the identity
\begin{equation}
-b^{2}x^{4}=-x^{2}(b^{2}x^{2}-1)-x^{2}~~~.
\end{equation}
Indeed
\begin{equation}
-\int_{0}^{1}\!\frac{b^{2}x^{4}{\rm d}x}
                 {(1-b^{2}x^{2})\sqrt{1-d^{2}-b^{2}x^{2}}}=
\int_{0}^{1}\!\frac{x^{2}{\rm d}x}{\sqrt{1-d^{2}-b^{2}x^{2}}}-
\int_{0}^{1}\!\frac{x^{2}{\rm d}x}{(1-b^{2}x^{2})\sqrt{1-d^{2}-b^{2}x^{2}}}
~~~.
\end{equation}
Thus
\begin{eqnarray}
\langle3\cos^{2}\theta_{1}\cos^{2}\theta_{2}\rangle
&=&\frac{C(g,s)}{d^{2}}
\left\{2\int_{0}^{1}\frac{x^{2}{\rm d}x}{\sqrt{1-d^{2}-b^{2}x^{2}}}-
\frac{1}{d}~{\rm arcsin}\frac{d}{\sqrt{1-b^{2}}}\right.\nonumber\\
&+&\left.\mbox{}\int_{0}^{1}\frac{x^{2}{\rm d}x}
                           {(1-b^{2}x^{2})\sqrt{1-d^{2}-b^{2}x^{2}}}\right\}
~~~,
\end{eqnarray}
which gives
\begin{eqnarray}
\langle3\cos^{2}\theta_{1}\cos^{2}\theta_{2}\rangle
&=&C(g,s)\left\{
-\frac{\sqrt{1-d^{2}-b^{2}}}{b^{2}}+
\frac{1-d^{2}}{b^{3}}~{\rm arcsin}\frac{b}{\sqrt{1-d^{2}}}\right.\nonumber\\
&-&\left.
\frac{1}{d}~{\rm arcsin}\frac{d}{\sqrt{1-b^{2}}}-
\frac{1}{b^{3}}~{\rm arcsin}\frac{b}{\sqrt{1-d^{2}}}+
\frac{1}{C(g,s)b^{2}}\right\}~~~.
\end{eqnarray}
Finally,
\begin{eqnarray}
\langle3\cos^{2}\theta_{1}\cos^{2}\theta_{2}\rangle
&=&C(g,s)\left\{
-\frac{\sqrt{1-d^{2}-b^{2}}}{d^{2}b^{2}}-
\frac{1}{b^{3}}~{\rm arcsin}\frac{b}{\sqrt{1-d^{2}}}\right.\nonumber\\
&-&\left.\frac{1}{d^{3}}~{\rm arcsin}\frac{d}{\sqrt{1-b^{2}}}+
\frac{1}{C(g,s)d^{2}b^{2}}\right\}~~~,
\end{eqnarray}
which is  symmetric in respect to the interchange $g\leftrightarrow 
s$ if both parameters are negative. The symmetry breaks if 
the parameters have opposite signs. Let
\begin{equation}
\begin{array}{ll}
s= b^{2} ~~~,\\
g=-d^{2} ~~~,\\
\end{array}
\end{equation}
then one has to calculate
\begin{eqnarray}
I_{3}
&=&C(g,s)\left\{
2\int_{0}^{1}\frac{x^{2}{\rm d}x}{\sqrt{1-d^{2}+b^{2}x^{2}}}+
\int_{0}^{1}\frac{x^{2}{\rm d}x}
                 {(1+b^{2}x^{2})\sqrt{1-d^{2}+b^{2}x^{2}}}\right.\nonumber\\
&-&\left.\frac{1}{d}~{\rm arcsin}\frac{d}{\sqrt{1+b^{2}}}\right\} 
~~~.
\end{eqnarray}
The result is
\begin{eqnarray}
I_{3}
&=&\frac{C(g,s)}{d^{2}}\left\{\frac{\sqrt{1-d^{2}+b^{2}}}{b^{2}}-
\frac{1-d^{2}}{b^{3}}~\ln
\left|\frac{b+\sqrt{1-d^{2}+b^{2}}}{\sqrt{1-d^{2}}}\right|\right.\nonumber\\
&+&\left.\frac{1}{b^{3}}~\ln\frac{b+\sqrt{1-d^{2}+b^{2}}}{\sqrt{1-d^{2}}}-
\frac{1}{b^{2}C(g,s)}-\frac{1}{d}~{\rm arcsin}\frac{d}{\sqrt{1+b^{2}}}\right\}
~~~.
\end{eqnarray}
If $s$ is negative while $g$ is positive, the integral is
\begin{eqnarray}
I_{3}
&=&\frac{C(g,s)}{b^{2}}\left\{-\frac{\sqrt{1+d^{2}-b^{2}}}{d^{2}}-
\frac{1-b^{2}}{d^{3}}~\ln
\left|\frac{d+\sqrt{1+d^{2}-b^{2}}}{\sqrt{1-d^{2}}}\right|\right.\nonumber\\
&+&\left.\frac{1}{d^{3}}~\ln\frac{d+\sqrt{1+d^{2}-b^{2}}}{\sqrt{1-d^{2}}}-
\frac{1}{d^{2}C(g,s)}-\frac{1}{b}~{\rm arcsin}\frac{b}{\sqrt{1+d^{2}}}\right\}
~~~.
\end{eqnarray}
For $s$ and $g$ both positive, one can obtain the integral by changing the 
inverse trigonometric functions with inverse hyperbolic functions:
\begin{eqnarray}
I_{3}
&=&C(g,s)\left\{-\frac{\sqrt{1+d^{2}+b^{2}}}{b^{2}d^{2}}+
\frac{1}{b^{3}}~{\rm arcsinh}\frac{b}{\sqrt{1+d^{2}}}\right.\nonumber\\
&+&\left.\frac{1}{d^{3}}~{\rm arcsinh}\frac{d}{\sqrt{1+b^{2}}}+
\frac{1}{C(g,s)d^{2}b^{2}}\right\} ~~~.
\end{eqnarray}
Note that this expression is also symmetric in respect to 
$s\leftrightarrow g$ interchange. 
The following formula valid for  all values of $s$ and $g$,
 sums up our results
\begin{equation}
\langle\cos^{2}\hat{\theta}\rangle=
\frac{1}{2gs}\left(1+g+s+gs-C(g,s)\sqrt{1+g+s}\right) ~~~.
\end{equation}
 Note, that $C(g,s)$
is given by Eq.~(\ref{cgs}) completely defines the solution for 
$\langle\cos^{2}\hat{\theta}\rangle$. 

\section{Analytics for perfect coupling}

Analytical solutions for the alignment measure corresponding to 
perfect coupling of $\bf J$ with
the axis of major inertia were obtained in Paper I. Here we write down
those solutions in the form convinient for comparing
with the solutions obtained in the main body of the present paper.

For oblate grains, 
\begin{equation}
\sigma=-\frac{3(1+g)}{2s}\left(1-\sqrt{-\frac{1+s+g}{s}}
\arcsin\sqrt{-\frac{s}{1+g}}\right)-\frac{1}{2}, ~~~s<0 ~~~.
\label{c1}
\end{equation}
and 
\begin{eqnarray}
\lefteqn{\sigma=-\frac{3(1+g)}{4s^{3/2}}
\left[2s^{1/2}
+(1+s+g)^{1/2}\right.}\nonumber\\
&\left.\mbox{}\times(\ln(1+g)-2\ln(s^{1/2}+(1+s+g)^{1/2}))\right]-\frac{1}{2}
, ~~~s>0 ~~~.
\label{c2}
\end{eqnarray}

\clearpage

\begin{center}
{\large \bf Figure Captions}\\
\end{center}

{\bf Fig.~1}. Distribution function of $J$ in the grain reference frame 
for $T_{\rm eff}/T_{\rm s}=100$\\ 

{\bf Fig.~2}. ${\rm Z}_{1}$ axis of the external or gas reference frame, 
${\rm X}_{1}{\rm Y}_{1}{\rm Z}_{1}$, is directed along magnetic field. The 
internal or grain frame, ${\rm X}_{2}{\rm Y}_{2}{\rm Z}_{2}$, is defined 
so that ${\rm Z}_{2}$ coincides with 
the symmetry axis of the spheroid. $\theta_{1}$, $\varphi_{1}$, $\theta_{2}$, 
and $\varphi_{2}$ are the polar angles in the above frames.\\

{\bf Fig.~3}. Rayleigh reduction factor, $R$, for oblate grains ($g<0$) 
subjected to a flux 
with $s<0$. A sharp peak corresponds to $g=s=-0.5$.\\

{\bf Fig.~4}. Rayleigh reduction factor, $R$, for oblate grains ($g<0$) 
under MHD waves ($s=-0.5$). The alignment is most efficient for 
flakes ($g=-0.5$). Due to the intrinsic symmetry inherent to the 
problem the same plot represents the Rayleigh reduction factor for flakes 
($g=-0.5$) when $s$ varies from $-0.5$ to 0 along the $x$-axis.\\

{\bf Fig.~5}. Rayleigh reduction factor, $R$, for flakes ($g=-0.5$) 
under streaming motions corresponding to  $s>0$. Evidently such an 
alignment is inefficient.\\

{\bf Fig.~6}. Rayleigh reduction factor, $R$, for oblate grains ($g<0$) when 
 $s>0$. The alignment measure for prolate 
grains when $s<0$ can be obtained simply by inchanging $s$ and $g$.\\

{\bf Fig.~7}. Rayleigh reduction factor, $R$, for prolate grains ($g>0$) if
${\bf u}\bot {\bf B}$ ($s=-0.5$). The alignment is marginal.\\

{\bf Fig.~8}.
Rayleigh reduction factor, $R$, for prolate grains ($g<0$) and $s>0$.
This situation corresponds to the alignment through streaming along
magnetic field lines.\\

{\bf Fig.~9}.
The measure of alinment, $R$, for prolate grains streaming along
magnetic field lines as a function of grain axis ratio $y$ and the
velocity ratio $w$. High degree of
alignment corresponds to both large $y$ and small $w$.

{\bf Fig.~10}.
Rayleigh reduction factor $R$ of prolate grains streaming along
magnetic field lines for $y=10$ (upper plot) and for $y=5$ (lower plot)
for a wide range of the velocities ratio $w$. It is evident, that for 
mildly supersonic drift the alignment is marginal.\\

{\bf Fig.~11}.
Rayleigh reduction factor $R$ of oblate grains subjected to the drift
with ${\bf u}\bot {\bf B}$ for $\bf J$ coupled with the axis of major inertia 
as a result of internal dissipation (upper plot) and for negligible 
internal dissipation (lower plot).\\

{\bf Fig.~12}. Rayleigh reduction factor of oblate grains subjected to 
drift with ${\bf u}\bot {\bf B}$. The solid line corresponds to the temperature
ratio 100, the dotted line corresponds to 10 and the dashed line corresponds
to 1.1. \\

{\bf Fig.~13}. Rayleigh reduction factor for  ${\bf u}\bot {\bf B}$
 as a funcion of 
grain eccentricity for prolate grains. The solid line corresponds to the 
temperature
ratio 100, the dotted line corresponds to 10 and the dashed line corresponds
to 1.1.\\

{\bf Fig.~14}. Rayleigh reduction factor for radiation pressure ($s=100$) 
as a funcion of 
grain eccentricity. The solid line corresponds to the temperature
ratio 100, the dotted line corresponds to 10 and the dashed line corresponds
to 1.1.

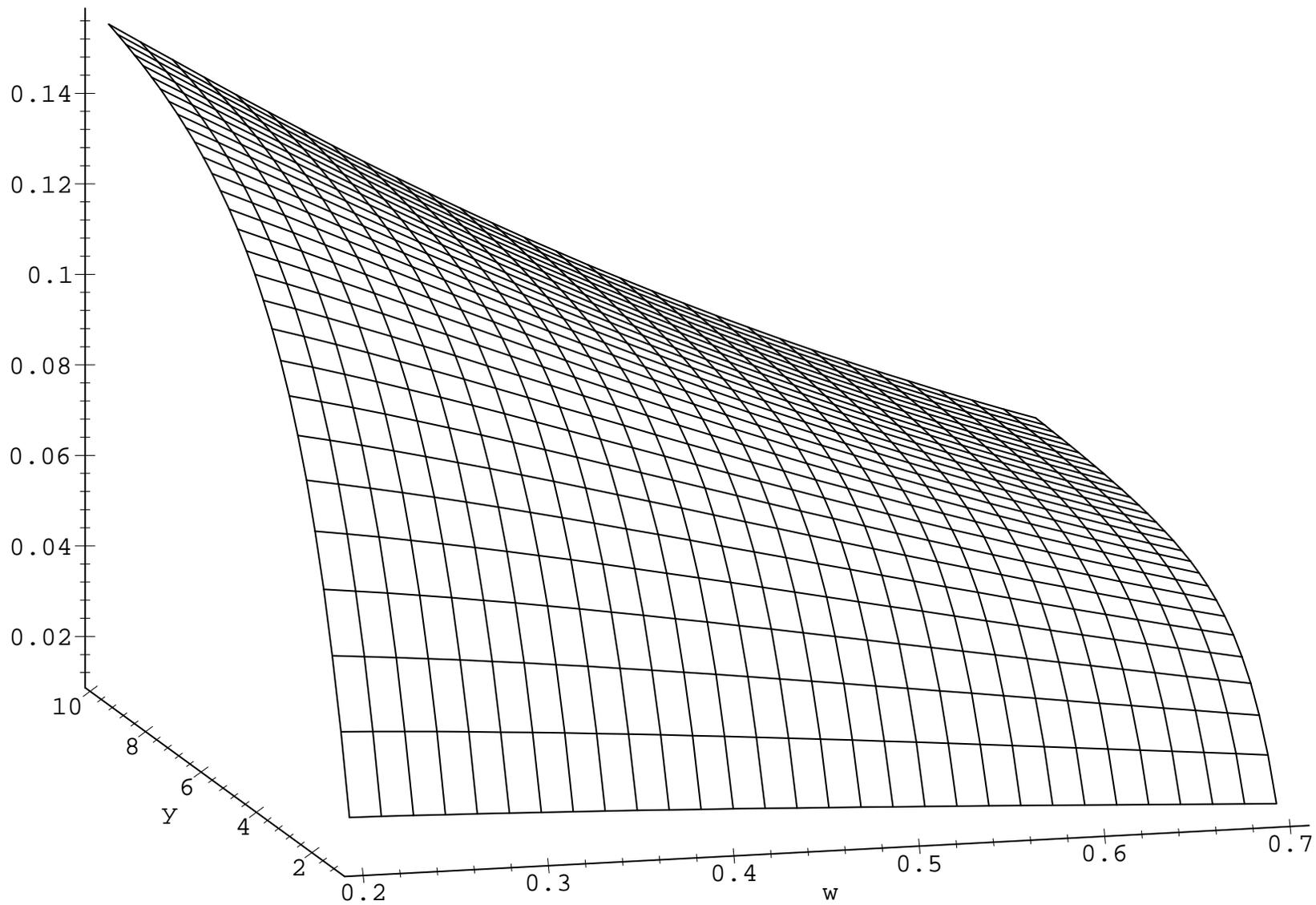
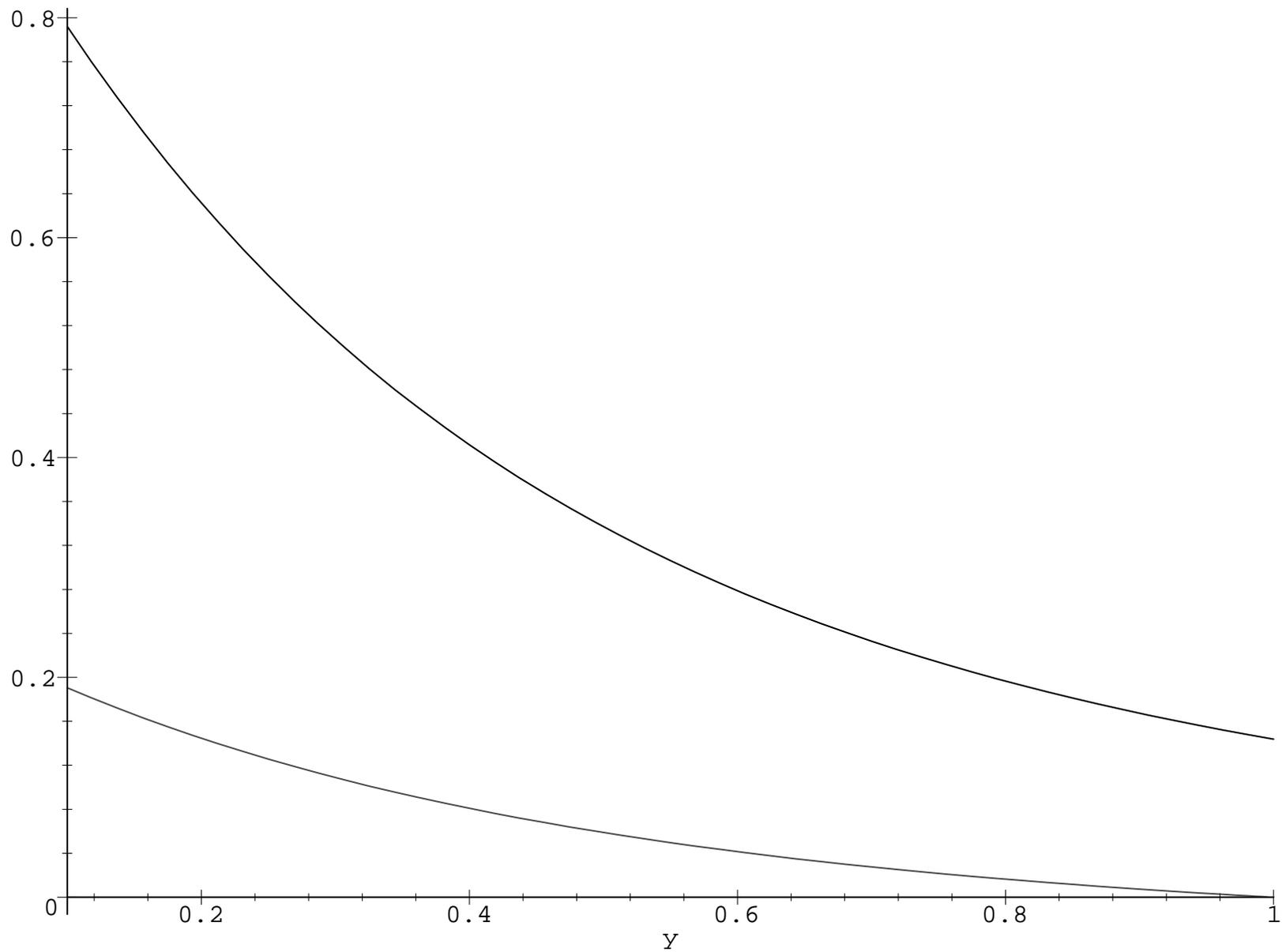
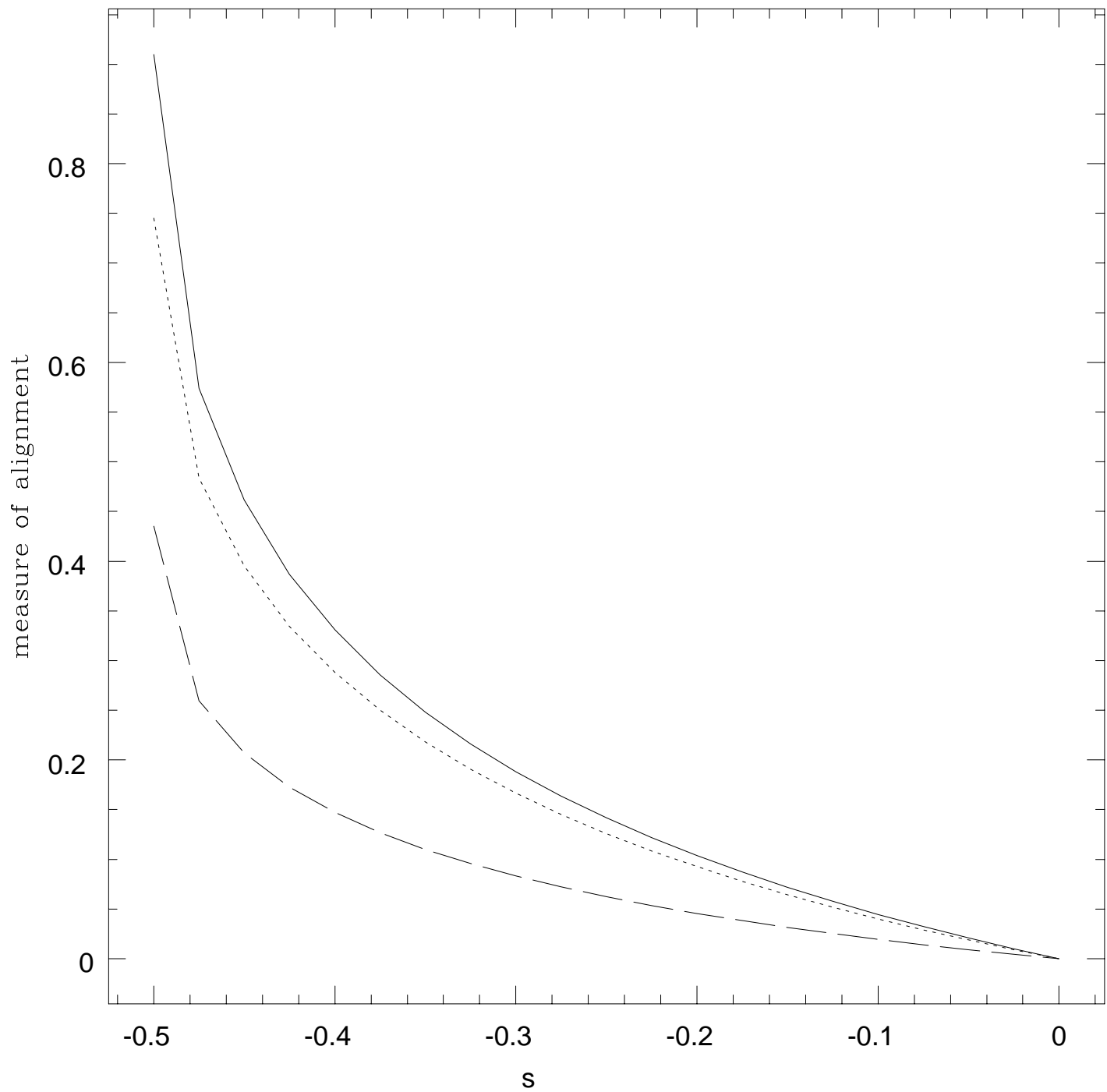
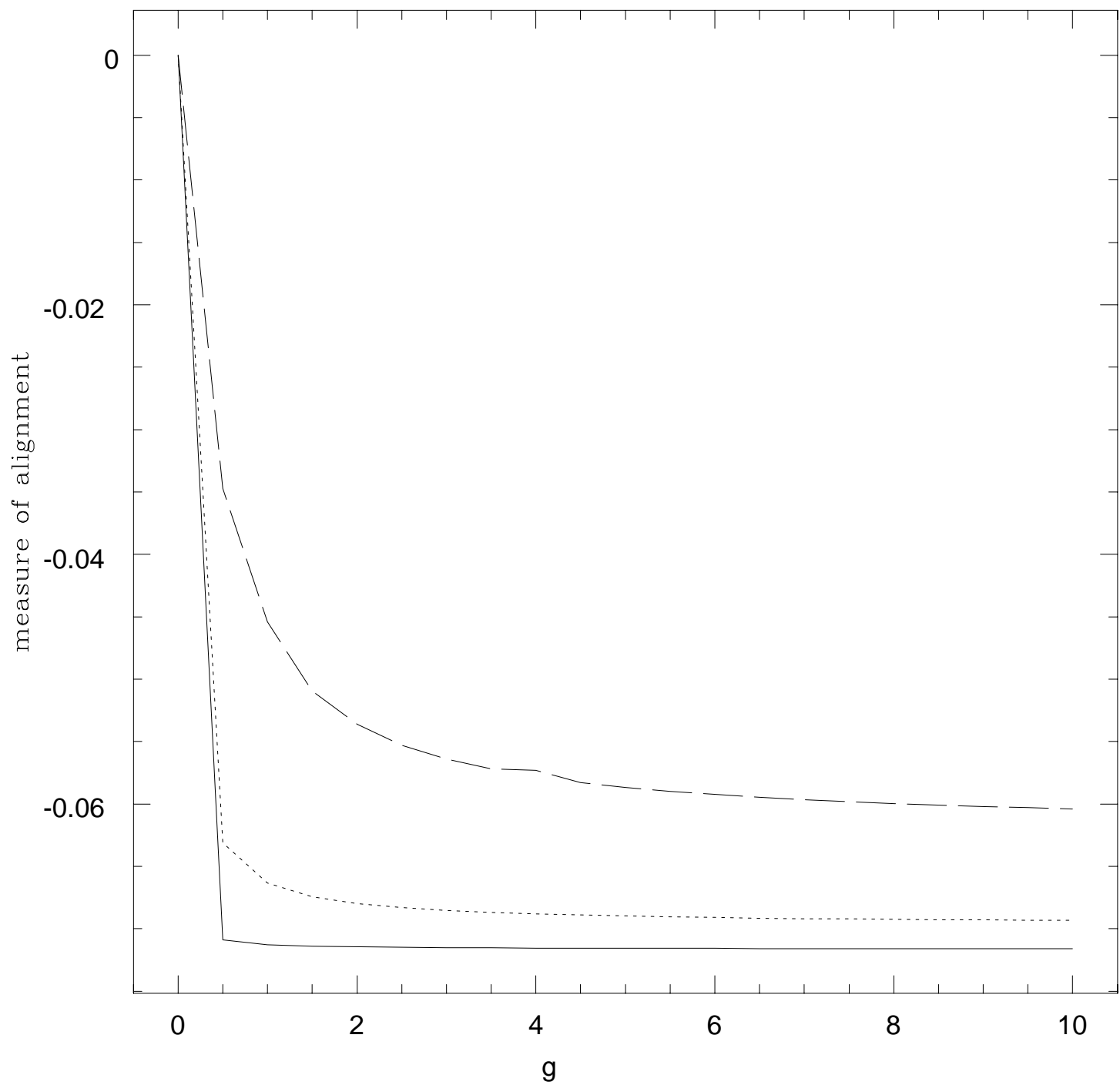
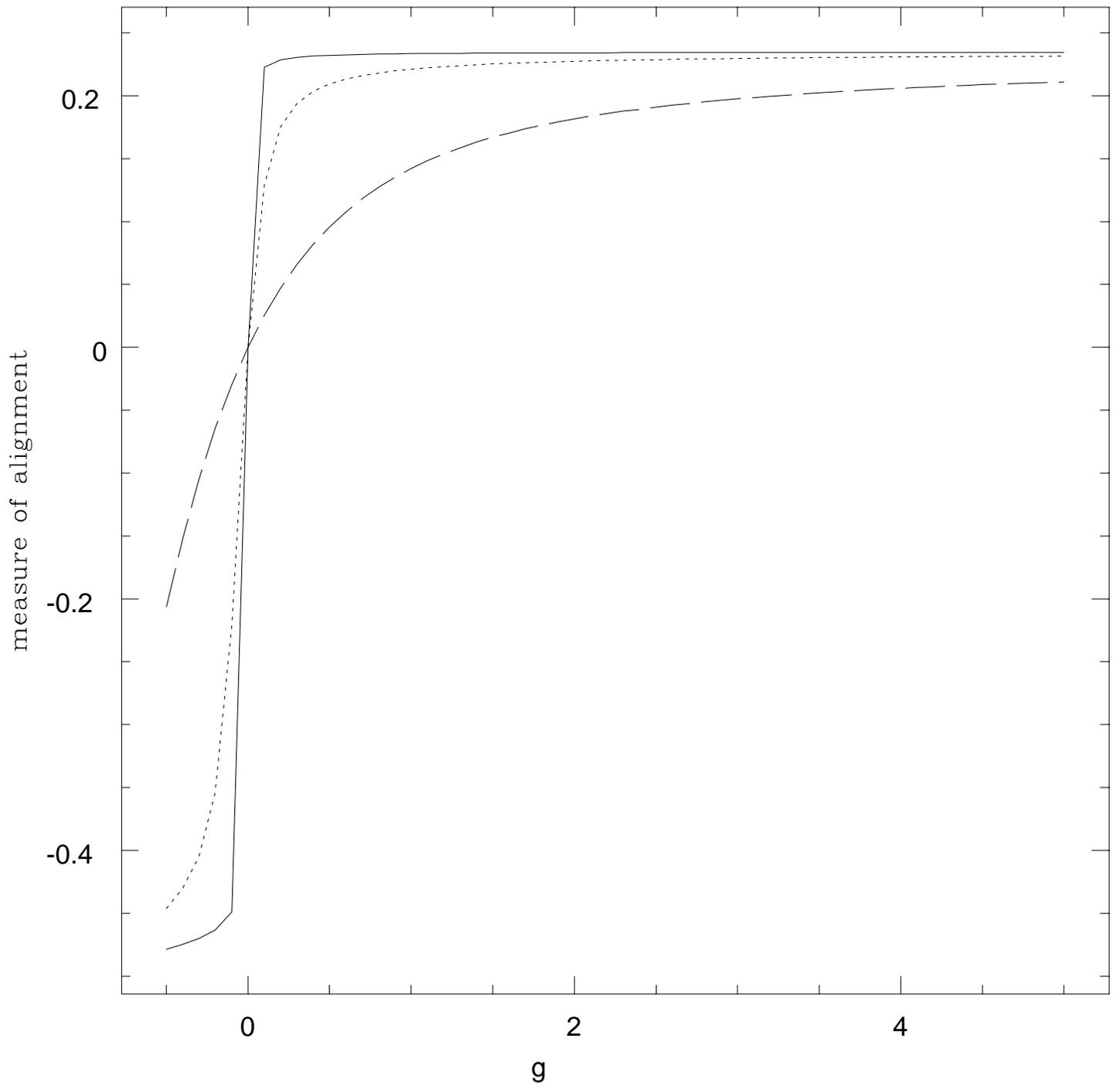
\end{document}